\documentclass[preprint,aps,prb]{revtex4}
\usepackage[dvips]{graphicx}
\usepackage{color,array,dcolumn}
\usepackage{amsmath}
\usepackage{amssymb}

\begin{document}

\title{van der Waals-corrected Density Functional Theory 
simulation of adsorption processes on transition-metal surfaces:
Xe and graphene on Ni(111)}

\author{Pier Luigi Silvestrelli and Alberto Ambrosetti}
\affiliation{Dipartimento di Fisica e Astronomia, 
Universit\`a di Padova, via Marzolo 8, I--35131, Padova, Italy,
and DEMOCRITOS National Simulation Center, of the Italian Istituto 
Officina dei Materiali (IOM) of the Italian National 
Research Council (CNR), Trieste, Italy}

\begin{abstract}
\date{\today}
The DFT/vdW-WF2s1 method, recently developed to include the van der Waals 
interactions in the Density Functional Theory and describe adsorption
processes on metal surfaces by taking metal-screening effects into 
account, is applied to the case of the interaction of Xe and graphene
with a transition-metal surface, namely Ni(111).
In general the adsorption of rare-gas atoms on metal surfaces is important 
because is prototypical for physisorption processes. 
Moreover, the interaction of graphene with Ni(111) 
is of particular interest for practical applications (efficient and large-scale
production of high-quality graphene) and, from a theoretical point of
view, is particularly challenging, since it can be described by
a delicate interplay between chemisorption and physisorption processes.
The first-principles simulation of transition metals require particular 
care also because they can be viewed as intermediate systems between 
simple metals and insulating crystals.
Even in these cases the method performs well
as demonstrated by comparing our results with available experimental 
data and other theoretical investigations.
We confirm that the rare gas Xe atom is preferentially adsorbed on the
top-site configuration on the Ni(111) surface too.
Our approach based on the use of the Maximally Localized Wannier Functions
also allow us to well characterize the bonds between graphene
and Ni(111). 
\end{abstract}

\maketitle

\section{Introduction}
The interaction of graphene with transition metal surfaces is of great
importance because the growth of graphene on these substrates is probably
at present the most perspective way for the large-scale preparation 
and production of high-quality graphene.\cite{Wintterlin,Voloshina2012}
In particular, the Ni(111) surface is very interesting, being the closest 
matched interface with respect to graphene of all transition metals 
(with a lattice mismatch of about 1\%), and has been therefore intensively 
investigated.\cite{Wintterlin,Voloshina2012,Batzill,Nagashima,Gamo,Fuentes,Zhao,DedkovPRL,Gruneis,Dedkov,Li,Allard,Riccardi,DedkovJAP,Cho,Odahara,Garcia,Bianchini,Zhang,Mittendorfer,Parreiras,Sun,Hamada,Olsen,Feng,Vanin,Andersen}
In fact, the close lattice match allows graphene to
adapt itself to the Ni(111) lattice and enables the formation of a 
well-ordered {\it {p($1 \times 1$)}} over-structure
on the Ni(111) surface, which is much simpler than the complex Moir\'e
patterns commonly found in other transition metal 
surfaces.\cite{Wintterlin,Batzill,Voloshina2012} Moreover, given the
ferromagnetic properties of Ni, graphene on Ni surface has also
been proposed as a promising spin-filtering device needed in 
spintronics.\cite{Karpan}
Due to the strong interaction 
between graphene $\pi$ and Ni $3d$ electrons, the graphene electronic 
structure is heavily modified if compared to the electronic structure of 
free-standing graphene. The most important change is the gap opening 
resulting from sublattice symmetry breaking.

From a fundamental point of view, the nature of the bonds at the 
graphene-metal interface has not been completely elucidated yet.
Basically, by taking the C-C covalent bonds as a reference, the 
interaction of graphene with a metal substrate can be
classified as ''weak'' (with Ir(111), Au(111), Cu(111), Pt(111), Al(111), 
and Pd(111)) or ''strong'' (with Ni(111), Co(0001), Rh(111), and 
Ru(0001)), see refs. \onlinecite{Batzill,Parreiras} and references therein.
Clearly this behavior depends on the atomic and electronic structures 
of the different substrates which can substantially modify the properties 
of the adsorbed graphene with respect to those of a free-standing
graphene layer. In particular,  
transition metals can be viewed as intermediate systems between simple
metals and insulating crystals because their conduction electrons are of
both free-electron ($s$- and $p$-like) character and of localized 
($d$-like) character, so that they are neither totally free nor completely 
bound. Moreover, the $d$-like electrons of the filled bands
are not rigidly bound to the ionic cores and therefore must be somehow
included in the description of screening effects.

From a theoretical point of view the interaction of graphene with Ni(111) 
represents a particularly challenging system, since,
both an accurate description of the metallic surface and of the
nonlocal correlation effects is needed.
Moreover, in the most favored adsorption
configurations, van der Waals (vdW) effects are 
very important although the graphene-metal interaction is relatively
strong and the equilibrium distance of about 2.1 \AA\ is much shorter 
than typical distances observed between fragments bonded by pure
vdW interactions, thus suggesting that a delicate interplay
between chemisorption and physisorption exists.\cite{Mittendorfer,Zhang}
As a result, the computed adsorption energies and equilibrium distances, 
even by adopting vdW-corrected Density Functional Theory (DFT) 
approaches, exhibit a disappointing wide spread of values.\cite{Zhang}
For instance, using the Generalized Gradient Approximation (GGA) 
functionals one has a very weak binding or no binding at all 
(see ref. \onlinecite{Zhang} and references therein), while 
the Local Density Approximation (LDA) calculations\cite{Zhang} lead to a 
pronounced minimum with an adsorption energy of about -200 meV per C atom, 
at an equilibrium distance of 2.00 \AA\ a little  
shorter than the experimental value of $2.11 \pm 0.07$ \AA.\cite{Gamo} 
Actually, the deep LDA minimum is a true artifact, being related to the 
wrong decay of the Kohn-Sham potential and orbitals from the surface into 
the vacuum. In any case
the relatively weak bonding confirms that the adsorption should 
not be viewed as a traditional (covalent) chemisorption characterized by 
typical adsorption energies of 0.5-2 eV, although the short graphene-Ni 
distances clearly indicate a covalent interaction already at the level of
semilocal GGA functionals.  

The adsorption of graphene on Ni(111) has been also investigated on the basis 
of the adiabatic-connection fluctuation-dissipation theorem in the 
random phase approximation (RPA).\cite{Mittendorfer,Olsen} 
At the moment this probably represents the more accurate calculation, which
can therefore be taken as a reasonable reference database, being
particularly suited to describe intricate bonds with mixed covalent and 
dispersive character,\cite{Mittendorfer,Olsen} although one should remember
that RPA typically exhibits a tendency to underbind.\cite{OlsenPRL,Ren,Harl}
In spite of a significant hybridization 
found between the graphene $\pi$ orbitals and Ni $d_{z^2}$
states at a binding distance of about 2.2 \AA, the computed RPA adsorption
energy is still in the range of a typical physisorption interaction
(about -70 meV per C atom).  
An important contribution to the energy is related to a decrease in 
the exchange energy resulting from the adsorption-induced lower 
symmetry in the graphene layer.
Interestingly, the RPA calculations do not only predict one
single minimum but also a second minimum at a distance of 3.3 \AA.
This second minimum, with an adsorption energy of -60 meV, that is
only slightly less bound than the most favored configuration, is found
at a distance typical for vdW-adsorbed graphene on 
transition metals.\cite{Sutter} 

This complex scenario has been rationalized as follows:\cite{Mittendorfer}
at large distances, namely up to 2.8 \AA, the electronic band structure of 
graphene on Ni(111) is hardly modified with respect to that of the 
free-standing, the exchange interaction is purely repulsive, and the 
correlation follows essentially a vdW-like behavior; instead, at distances 
shorter than 2.8 \AA, the 
graphene band structure starts to be modified with a hybridization setting 
in, and, most importantly, with a breakup of the symmetry between the 
{\it{top}}-site 
and {\it{hollow}}-site C atoms. As a delicate balance between exchange
and correlation energy the total energy is smooth, with a slight barrier 
between the physisorption minimum, characterized by a graphene band structure 
that is hardly modified compared to free-standing graphene, and another 
''chemisorption'' minimum, where the graphene band structure is 
strongly modified compared to the free-standing layer.
The physisorption minimum originates entirely from vdW forces, whereas the 
chemisorbed minimum can be attributed to both orbital hybridization 
and week vdW forces. The slight energy barrier between physical adsorption 
and chemisorption state implies that the graphene can be decoupled easily 
from Ni(111) surface, which is also evidenced in chemical vapor deposition 
(CVD) growth of graphene on Ni surface.\cite{Zhang} 
Besides vdW effects at large distances it is therefore also important to 
describe exchange interaction at a short distance where the correct 
description of Pauli repulsion and hybridization is crucial.\cite{Hamada} 

Since graphene has two carbon atoms in the unit cell,
while the Ni(111) surface has only one, six different structural models are 
possible, namely {\it{top-fcc}}, {\it{top-hcp}}, {\it{bridge-top}}, {\it{fcc-hcp}}, {\it{bridge-fcc}}, and {\it{bridge-hcp}}. 
In the {\it{top-fcc}} and {\it{top-hcp}} models, one C atom sits on top of 
the Ni atom while the other C atom occupies, respectively, the fcc and hcp 
hollow sites. The {\it{bridge-top}} configuration has the carbon-carbon bond of 
graphene on top of a surface Ni atom. In the {\it{fcc-hcp}} model, both C atoms 
are in the 3-fold hollow sites (see Fig. 1).

Recently, the structural properties of graphene/Ni(111) have been 
experimentally investigated\cite{Parreiras} by a combination of 
low-energy electron diffraction (LEED), x-ray photoelectron spectroscopy 
(XPS), and angle-scanned photoelectron diffraction (PED).
XPS data indicate that graphene interacts strongly with the topmost Ni 
layer and diffraction data show that graphene is deposited commensurably with 
the underlying Ni surface atoms. Interestingly, related
first-principles calculations based on DFT show\cite{Parreiras} that the 
total energies of the {\it{top-fcc}} 
and {\it{bridge-top}} structures are nearly degenerate, which corroborates the 
observed\cite{Bianchini,Parreiras} coexistence of those phases. 
A depletion of charge density in Ni-$d_{z^2}$ of
topmost Ni atoms is also found and an increase of that in 
$p_z$ of on-top C
atom in the {\it{top-fcc}} and {\it{top-hcp}} structures, which indicates 
that the electrons 
transfer from substrate to adsorbed graphene.\cite{Zhang}

From a theoretical point of view, it is well accepted that
the widely-used LDA and semi-local GGA functionals fail to capture vdW 
forces, which are important in rare-gas/metal and graphene/metal systems. 
As mentioned above, 
the most reliable RPA calculations\cite{Mittendorfer,Olsen} 
show that the binding in graphene/Ni system is a delicate 
balance between covalent and dispersive interactions. However, the RPA 
calculation is computationally expensive (scaling as $N^4$ with system size)
and thus hardly accessible for large systems. 
The cheaper DFT-D approach\cite{PBE+D} with semi-empirical corrections for 
vdW interactions seems to typically give a reasonable prediction of the 
adsorption properties. However, its empirical 
character\cite{Zhao,Kozlov,Sun11,Hasegawa}
casts doubts on the full reliability and accuracy in many applications,
particularly in metal system where the atom-based description, implied in the
use of the $C_6$ coefficients adopted in DFT-D, is rather questionable.

In the last few years several practical methods have been proposed
to make DFT calculations able to accurately describe vdW effects at a 
reasonable computational cost (for a
recent review, see, for instance, refs. \onlinecite{Riley,MRS,Klimes}).
We have developed a family of such methods, all based on the generation
of the Maximally Localized Wannier Functions (MLWFs),\cite{Marzari}
successfully applied to a variety of 
systems, including small molecules, water clusters, 
graphite and graphene, water layers interacting with graphite, 
interfacial water on semiconducting substrates,
hydrogenated carbon nanotubes, 
molecular solids, and the interaction of rare gases and small molecules 
with metal surfaces.\cite{silvprl,silvmetodo,silvsurf,CPL,silvinter,ambrosetti,Costanzo,Ar-Pb,Ambrosetti2012,C3,PRB2013,QHO-WF,QHO-WFs}
In particular, the DFT/vdW-WF2s1 method, presented in 
ref. \onlinecite{PRB2013}, has been specifically
developed to take metal-screening effects into account 
and has been applied to the study of the
adsorption of rare gases and small molecules on different metal surfaces,
namely Al(111), Cu(111), and Pb(111).

Here we apply the DFT/vdW-WF2s1 approach to investigate the interaction of
the Xe atom and of graphene with the Ni(111) surface.
Our results will be compared to the best available, reference experimental
and theoretical values, and to those obtained by other DFT
vdW-corrected schemes, including PBE+D\cite{PBE+D} (a particular DFT+D scheme), 
vdW-DF,\cite{Dion,Langreth07} vdW-DF2,\cite{Lee-bis} rVV10,\cite{Sabatini}
and by the simpler Local Density Approximation (LDA) and semilocal GGA
(in the PBE flavor\cite{PBE}) approaches.
In the PBE+D scheme DFT calculations at the PBE level are corrected
by adding empirical $C_6/R^6$ potentials with parameters determined by fitting
accurate energies for a large molecular database, while in other methods,
such as vdW-DF, vdW-DF2, and rVV10, vdW effects are included by 
introducing DFT nonlocal correlation functionals.

\section{Method}
Here we briefly review the DFT/vdW-WF2s1 method; additional details can be
found in refs. \onlinecite{C3,PRB2013}.
Basically, the scheme relies on the    
well known London's expression\cite{london} where
two interacting atoms, $A$ and $B$, are approximated by 
coupled harmonic oscillators
and the vdW energy is taken to be the change of the zero-point energy
of the coupled oscillations as the atoms approach; if only a single
excitation frequency is associated to each atom, $\omega_A$, $\omega_B$,
then

\begin{equation}
E^{London}_{vdW}=-\frac{3e^4}{2m^2}\frac{Z_A Z_B}{\omega_A \omega_B(\omega_A+\omega_B)}\frac{1}{R_{AB}^6}
\label{lond}
\end{equation}

where $Z_{A,B}$ is the total charge of A and B, and $R_{AB}$ is 
the distance between the two atoms ($e$ and $m$ are the electronic charge
and mass).

Now, adopting a simple classical theory of the atomic polarizability, 
the polarizability 
of an electronic shell of charge $eZ_i$ and mass $mZ_i$, tied to a heavy 
undeformable ion can be written as

\begin{equation}
\alpha_i\simeq \frac{Z_i e^2}{m\omega_i^2}\,.
\label{alfa}
\end{equation}

Then, given the direct relation between polarizability and atomic
volume,\cite{polvol} we assume that $\alpha_i = \gamma S_i^3$,
where $\gamma$ is a proportionality constant, so that the atomic volume is
expressed in terms of the MLWF spread, $S_i$.
Rewriting eq. \eqref{lond} in terms of the quantities defined above,
one obtains an explicit expression for the $C_6$ vdW coefficient:

\begin{equation}
C_{6}^{AB}=\frac{3}{2}\frac{\sqrt{Z_A Z_B}S_A^3 S_B^3 \gamma^{3/2}}
{(\sqrt{Z_B}S_A^{3/2}+\sqrt{Z_A}S_B^{3/2})}\,.
\label{c6}
\end{equation}

The constant $\gamma$ can then be set up by imposing that the exact value for
the H atom polarizability
($\alpha_H=$4.5 a.u.) is obtained (of course, in the H case, one
knows the exact analytical spread, $S_i=S_H=\sqrt{3}$ a.u.).

In order to achieve a better accuracy, one must properly deal with
{\it intrafragment} MLWF overlap (we refer here to charge overlap, not to be
confused with wave functions overlap). 
This overlap affects the effective orbital volume,
the polarizability, and the excitation
frequency (see eq. \eqref{alfa}), thus leading to a quantitative effect on the
value of the $C_6$ coefficient.
We take into account the effective change in volume due to intrafragment
MLWF overlap by introducing a suitable reduction factor $\xi$
obtained by interpolating between the limiting cases of fully
overlapping and non-overlapping MLWFs (see ref. \onlinecite{C3}).
We therefore arrive at the following expression for the $C_6$ coefficient:

\begin{equation}
C_{6}^{AB}=\frac{3}{2}\frac{\sqrt{Z_A Z_B}\xi_A S_A^3 \xi_B S_B^3 \gamma^{3/2}}
{(\sqrt{Z_B\xi_A} S_A^{3/2}+\sqrt{Z_A\xi_B} S_B^{3/2})}\,,
\label{c6eff}
\end{equation}

where $\xi_{A,B}$ represents the ratio between the effective and the
free volume associated to the $A$-th and $B$-th MLWF.

Finally, the vdW interaction energy is computed as:

\begin{equation}
E_{vdW}=-\sum_{i<j}f(R_{ij})\frac{C_6^{ij}}{R^6_{ij}}
\label{EvdW}
\end{equation}
where $f(R_{ij})$ is a short-range damping function, which is
introduced not only to avoid the unphysical divergence of the
vdW correction at small fragment separations, but also
to eliminate double countings of correlation effects
(in fact standard DFT approaches are able to describe short-range
correlations); it is defined as:
\begin{equation}
f(R_{ij})=\frac{1}{1+e^{-a(R_{ij}/R_s-1)}}\,.
\end{equation}

The parameter $R_s$ represents
the sum of the vdW radii $R_s=R_i^{vdW}+R_j^{vdW}$, with
(by adopting the same criterion chosen above for
the $\gamma$ parameter)
\begin{equation}
R_i^{vdW}=R_H^{vdW}\frac{S_i}{\sqrt{3}}
\end{equation}
where $R_H^{vdW}$ is the literature\cite{Bondi} (1.20 \AA) vdW radius of
the H atom, and, following Grimme {\it et al.},\cite{Grimme}
$a \simeq 20$ (the results are almost
independent on the particular value of this parameter).
Although this damping function introduces a certain degree of empiricism
in the method, we stress that $a$
is the only ad-hoc parameter present in our approach, while all the others
are only determined by the basic information given by the MLWFs, namely
from first principles calculations.

To get an appropriate inclusion of metal screening effects
a proper reduction coefficient is included by multiplying the
$C_6^{ij}/R^6_{ij}$ contribution in eq. \eqref{EvdW} by a Thomas-Fermi factor:
$ f_{TF} = e^{-2(z_s-z_l)/r_{_{TF}}}$
where $r_{_{TF}}$ is the Thomas-Fermi screening length relative
to the electronic density of an effective uniform electron 
gas (''jellium model'') describing the substrate (see discussion below),
$z_s$ is the average vertical position of the topmost metal
atoms, and $z_l$ is the vertical coordinate of the Wannier Function Center
(WFC) belonging to the
substrate ($l=i$ if it is the $i$-th WFC which belongs to the
substrate, otherwise $l=j$); the above $f_{TF}$ function is only applied if
$z_l < z_s$, otherwise it is assumed that $f_{TF}=1$ (no screening effect). 

\subsection{Computational details}
We here apply the DFT/vdW-WF2s1 method to the case of adsorption of 
Xe and graphene on the Ni(111) surface.
All calculations have been performed 
with the Quantum-ESPRESSO ab initio package\cite{ESPRESSO}
(MLWFs have been generated as a post-processing calculation using
the WanT package\cite{WanT}). Similarly to our previous 
studies\cite{Ar-Pb,PRB2013} we modeled the metal surface 
using a periodically-repeated hexagonal supercell, 
with a $(\sqrt{3}\times \sqrt{3})R30^{\circ}$ structure and a surface slab
made of 15 Ni atoms distributed over 5 layers
(repeated slabs were
separated along the direction orthogonal to the surface by a vacuum region
of about 25 \AA\ to avoid significant spurious interactions due to periodic
replicas), considering the experimental Ni(111) lattice
constants. The Brillouin Zone has been sampled using a 
$6\times6\times1$ $k$-point mesh. 
Above one of the surface Ni layers we add a Xe atom or a single 
graphene layer. We remind that
the Ni(111) surface and the graphene lattice constants are, respectively, 
2.489 and 2.46 \AA, which corresponds to a lattice misfit of about 1.2\%,
so that one can assume that the graphene layer lattice vectors
are commensurate with those of the Ni(111) surface. 
In this model system, with a Xe atom per supercell, 
the coverage is 1/3, i.e. one adsorbed adparticle for each 3
metal atoms in the topmost surface layer. 
The $(\sqrt{3}\times \sqrt{3})R30^{\circ}$
structure has been indeed observed\cite{Seyller} at low temperature by LEED for
the case of Xe adsorption on Cu(111) and Pd(111) (actually, 
this is the simplest commensurate structure for rare-gas monolayers on 
close-packed metal surfaces and the only one for which good 
experimental data exist), and it was adopted in most of 
the previous ab initio 
studies.\cite{Silva,DaSilva05,DaSilva,Righi,Lazic,Zhang11}
The metal surface atoms
were kept frozen (of course after a preliminary relaxation of the outermost
layers of the clean metal surfaces) and only the 
vertical coordinate (perpendicular to the surface) 
of the Xe atom or the graphene layer
was optimized, this procedure being justified
by the fact that only minor metal surface atom displacements are 
observed,\cite{DaSilva05,Zhang11,Abad,Fajin} and relaxation effects
are estimated to be small.\cite{Hamada12}
Also in the case of graphene, which seems to chemisorb on Ni(111), the
buckling between nonequivalent carbon atoms is negligible, indicating that
the adsorbed graphene layer is quite smooth.\cite{Sun}
Given the ferromagnetic character of Ni, spin polarization was 
taken into account.

We have carried out calculations for various separations of the Xe atom 
and the graphene layer, with Xe adsorbed on the {\it top} and {\it hollow}
(on the center of the triangle formed by the 3 surface
metal atoms contained in the supercell)
high-symmetry sites, and considering the {\it{top-fcc}}, 
{\it{bridge-top}}, and {\it{fcc-hcp}}
configurations for graphene on the metal substrates (see above).  
For a better accuracy, as done in previous applications on adsorption
processes,\cite{silvsurf,silvmetodo,silvinter,ambrosetti,Ar-Pb,PRB2013} 
we have also
included the interactions of the MLWFs of the adsorbate not 
only with the MLWFs of the underlying surface, within the reference supercell,
but also with a sufficient
number of periodically-repeated surface MLWFs (in any case, given the
$R^{-6}$ decay of the vdW interactions, the convergence with
the number of repeated images is rapidly achieved).
Electron-ion interactions were described using norm-conserving
pseudopotentials by explicitly including
10 valence electrons per Ni atom.
We chose the PBE\cite{PBE} reference DFT functional, which is probably
the most popular GGA functional. 

The choice of a suitable effective value for the uniform electron density $n$
(or, equivalently, the dimensionless ''$r_s$'' parameter, 
$r_s = \left(3/{4\pi n}\right)^{1/3}/{a_0}$, 
where $a_0$ is the Bohr radius) associated to the metal substrate to get the 
Thomas-Fermi length $r_{_{TF}}$, defined above to describe screening
effects, deserves a specific comment since we are considering 
transition metals where, differently form simple metals, the electrons
cannot be assumed to be totally free.
To address this issue we used the list of values for the effective 
free-electron density parameter $r_s$ provided by Perrot and 
Rasolt\cite{Perrot} for transition metals.
These values were obtained by considering the problem of
appropriately describing the ground state properties (particularly
those of arbitrary defects) of the
mobile part of the electron fluid in the transition metals and
defining the response of the ''free'' mobile part of the electron fluid using
an effective $r_s$ of a uniform electron gas.
The recipe is based on the concept of the metallic response to external 
low-symmetry perturbations and the obtained picture is that,
as expected, the $s$ and $p$ electrons carry the main response to an 
external potential while the $d$ electrons remain largely unpolarized.
From the $r_s$ parameter $r_{_{TF}}$ can be evaluated using the following 
formulas:\cite{Ashcroft}

\begin{equation}
k_F =\left(\frac{9\pi}{4}\right)^{1/3} \frac{1}{r_s a_0} \; ,
\end{equation}

\begin{equation}
k_{TF}^2 = 4\pi e^2 g({\epsilon}_F) = \frac{4 m e^2}{\pi \hbar^2} k_F \; ,
\end{equation}

\begin{equation}
r_{TF} = 1/k_{TF} \simeq \frac{\sqrt{r_s}}{1.563} a_0 \; ,
\end{equation}

where $k_F$ and $k_{TF}$ are the Fermi and Thomas-Fermi wavevector, 
respectively, and $g({\epsilon}_F)=m k_F/{\hbar^2 \pi^2}$ is the 
density of levels at the Fermi energy ${\epsilon}_F$ in the jellium model.
Using the $r_s = 2.14$ value suggested for Ni by 
Perrot and Rasolt,\cite{Perrot} one obtains a Thomas-Fermi screening length
$r_{_{TF}}= 0.496$ \AA.

\section{Results and Discussion}
In Table I we report the binding energy and the equilibrium
distance for Xe on Ni(111).
Note that all the methods predict that the {\it top} configuration is 
favored with respect to the {\it hollow} one,
in line with the observed general tendency of Xe and Kr for
adsorption on metallic surfaces in the low-coordination {\it top} 
sites.\cite{Diehl,DaSilva05,DaSilva,Betancourt,Lazic} This behavior
is attributed\cite{Diehl,Bagus} to the delocalization
of charge density that increases the Pauli repulsion effect at the
{\it hollow} sites relative to the {\it top} site and lifts the potential well
upwards both in energy and height.

The experimentally measured {\it adsorption energy}, $E_a$, often
includes not only the interaction of adparticles with the substrate but
also lateral, vdW interfragment interactions.\cite{Sun11,Ar-Pb}
Therefore sometimes it is more appropriate to compare experimental
data with the quantity $E_a$ which can be related to $E_b$ by:\cite{Ar-Pb}
                                                                         
\begin{equation}
E_a = E_b + (E_l - E_f) \;,
\label{Ea}
\end{equation}
                                                                        
where $E_l$ is the total energy (per particle) of the 2D lattice formed by the
adparticles only
(that is as in the adsorption configurations but without the substrate
and including vdW interfragment corrections when vdW-corrected
methods are used) and $E_f$ is the energy of an isolated (free)
adparticle.

Experimental estimates of the adsorption energy for Xe on Ni(111) range
from -370 to -180 meV.\cite{Vidali,Dolle,Wong} 
These data were obtained from measurements of the isosteric heat of 
adsorption\cite{Dolle} and optical differential reflectance studies
of adsorption and desorption of Xe on Ni(111).\cite{Wong}
In the latter investigations the overlayer structure at the 
saturation coverage (1 Xe atom per 3
top Ni atoms) appears commensurate with the underlying Ni(111) lattice,
which is consistent with the fact that the Xe-Xe separation on Ni(111)
is 4.31 \AA, which is compressed from the bulk Xe equilibrium value of
4.34 \AA\ by only 0.7\%. 

Our computed binding energy of Xe on Ni(111) is similar to that 
obtained by the rVV10 functional, while it is smaller (in absolute
value) than the values predicted by LDA and PBE+D and larger than
the vdW-DF2 value (and obviously the PBE one since this functional
does not properly include vdW interactions). 
Interestingly, 
by taking the Xe-Xe interaction into account to estimate $E_a$, our 
DFT/vdW-WF2s1 value 
is compatible with the experimental estimates, while instead LDA and PBE+D
appear to significantly overbind. 
To make the comparison with the experiment more accurate one must point out
that, although Xe on Ni(111) is considered a typical case of physisorption
process, a net transfer of electronic charge is observed from the
Xe atom to the metal, which leads to a surface-induced weakening
of the Xe-Xe interaction resulting from the Coulomb repulsion due
to the charging of the adatoms.\cite{Muller}
As a result, the absolute value of the correction factor 
$|(E_l - E_f)| = 88 $ meV
in eq. (\ref{Ea}) is probably overestimated; in fact, according to 
ref. \onlinecite{Wong}, the difference between $E_a$ and $E_b$
should be in the range of 30-40 meV. Considering this 
further correction, our DFT/vdW-WF2s1 estimate for $E_a$ would be
about -320 eV, always in line with the experimental values. 

In Table II we report our computed binding energies and
equilibrium distances for graphene on Ni(111), compared
to data obtained by other methods 
and literature experimental and theoretical estimates 
(in this case much more data are available than for Xe on Ni(111)).
The binding energy curves for graphene on Ni(111), 
obtained using our DFT/vdW-WF2s1 method and relative to the three
considered configurations, are plotted in Fig. 2.

By focusing on the {\it{top-fcc}} configuration, as discussed above, 
although not necessarily exact,\cite{OlsenPRL} the RPA 
estimate\cite{Mittendorfer,Olsen,Zhang} can be assumed 
to be the ''best'' theoretical estimate to be taken as a meaningful reference.
As expected, the PBE functional, which does not include genuine vdW effects,
dramatically underestimates the binding so that the binding energy is 
positive which means that the system is actually unbound, although the
equilibrium distance, corresponding to a local minimum in the binding energy, 
agrees well with the reference estimates.
However, even the performances of some vdW-corrected functionals are not
satisfactory by stressing the difficulty of properly describing this
system. In fact, for instance, rVV10 largely overestimate the equilibrium 
distance and, by comparison with the RPA curve, it appears to completely 
miss the first (chemisorption) minimum.
The vdW-DF and vdW-DF2 methods underestimate the binding energy
but (above all) predict a much too large equilibrium distance.
Other approaches (see Table II) give reasonable equilibrium distances,
although the estimated binding energies are significantly scattered,
with the optB86b\cite{Mittendorfer} and optB88vdW\cite{Zhang} methods 
which appear to give the best performances. 

For a better assessment of the different approaches, one should 
remember that RPA tends to underbind:\cite{OlsenPRL,Ren,Harl} for instance,
in the case of the interaction of CO with the Cu(111) surface,
RPA remarkably predicts the correct favored adsorption site, however
it underestimates\cite{Harl} the adsorption energy by 70 meV, corresponding
to a significant error of about 15\% with respect to the both the experimental
reference value and high-level quantum chemistry (CASPT2) calculations. 
By taking this observation into account and 
looking at both the binding energy and the equilibrium distance, 
the performances of DFT/vdW-WF2s1 are rather good, being also better
that those of the semiempirical DFT-D2 and DFT-D3 approaches:
in fact, DFT-D2 tends to overbind, while DFT-D3 evidently overestimates
the equilibrium distance.
Interestingly, the DFT/vdW-WF2s1 results are similar to those obtained by
the PBEsol functional, but for the {\it{fcc-hcp}} configuration.
This behavior is understandable since PBEsol is a semilocal GGA 
functional\cite{PBEsol} (a revision of PBE to better describe 
solid state and surface systems) which does not properly include
vdW interactions and therefore cannot reproduce the energy minimum
of the {\it{fcc-hcp}} structure that is essentially determined by long-range 
vdW effects. 
Our DFT/vdW-WF2s1 results indicate that the {\it{top-fcc}} configuration is 
only marginally more favored than the {\it{bridge-top}}, 
being in line with recent
experimental measurements,\cite{Bianchini,Parreiras} which suggest that the  
total energies of the {\it{top-fcc}}
and {\it{bridge-top}} structures are nearly degenerate since the two phases are 
observed to coexist.
Note that LDA and PBEsol, which do not take vdW effects into account,
differently from the vdW-corrected DFT/vdW-WF2s1, DFT-D2, and rVV10
methods, indicate the {\it{bridge-top}} as energetically more favored than
{\it{top-fcc}}; although this result should not be overemphasized, given the 
relatively small energy differences between the two configurations, 
nonetheless it suggests that a proper inclusion of vdW interactions
can be of importance even when covalent interactions are involved,
as already observed elsewhere.\cite{Liu}
Interestingly, the binding-energy curve of the {\it{bridge-top}} configuration
in Fig. 2 exhibit a shallow minimum at about 3.2 \AA\, 
in addition to the first,
more pronounced minimum at shorter distance, in qualitative agreement 
with the RPA findings\cite{Mittendorfer,Olsen,Zhang} 
(for the {\it{top-fcc}} configuration this second minimum is not 
clearly visible).

In Fig. 2, 
for the {\it{top-fcc}} case only, we also report the 
binding-energy curve obtained using the DFT/vdW-WF2s1 scheme where,
however, at each graphene-Ni(111) distance, the MLWFs used for the
calculation of the vdW energy are just those obtained at the largest 
distance (5.13 \AA) but rigidly shifted along the vertical 
$z$ coordinate in such a way to be placed
at the same $z$ level of the graphene layer (DFT/vdW-WF2s1-shift scheme).
As can be seen, this curve is very different from that obtained by 
considering the MLWFs consistently generated at each graphene-Ni(111) distance;
in particular it predicts that the equilibrium value (see Table II) is
at a much larger distance, as expected for a configuration dominated by
standard long-range vdW interactions.
This clearly demonstrates that the rearrangement of the MLWFs
when graphene approaches the Ni(111) substrate, as a consequence of 
the bond formation, is indeed crucial for a reasonable description of the
energetics of the system. 
This observation is confirmed by analyzing Fig. 3 where 
histograms are plotted showing the distribution of the WFCs along the
$z$ coordinate (perpendicular to the Ni(111) surface), by also
taking the relative spreads of the MLWFs into account as a Gaussian
smearing.
By considering, for the {\it{top-fcc}} configuration, two different 
graphene-Ni(111) distances, representative for the chemisorbed structure
(corresponding to the binding energy minimum) and for the 
physisorbed structure (corresponding to the largest distance), respectively,
the different distribution is evident.
In fact, at large graphene-Ni(111) distance there is a clear separation
between the WFCs "belonging" to the graphene layer and those describing
the electronic charge distribution of the Ni(111) substrate 
(mainly localized at positions along $z$ corresponding to the locations of the
metal layers), while, at the shorter
equilibrium distance, 3 WFCs, corresponding to MLWFs characterized by
larger spreads (about 1.6 \AA) than those of typical MLWFs of free-standing
graphene (about 0.8 \AA), are below graphene at an intermediate position 
between graphene and Ni(111). By also considering (see Fig. 1) that 3 is just
the number of C atoms located on top positions and thus able to form
covalent bonds with the Ni substrate, within the simulation cell,   
this feature well illustrates the chemical bonding formation even without the
need of a detailed analysis of the electron density distribution.   
Also note that the number of WFCs localized slightly above the topmost
Ni layer increases from 2 to 6 as the graphene-Ni(111) distance is decreased
from 5.13 to 2.12 \AA, in line with the reported\cite{Zhang} tendency 
of an electron transfer from the Ni(111) substrate to adsorbed graphene.
Moreover, the dashed curve obtained by the Gaussian 
smearing clearly suggests that a significant overlap between the MLFWs
belonging to graphene and those of the Ni(111) substrate occurs.  
Such an overlap is instead absent at the largest distance where evidently no
chemical bond between graphene and Ni(111) is formed.

\section{Conclusions}
In conclusion, we have extended the applicability of our DFT/vdW-WF2s1 method 
by investigating the interaction of Xe and graphene with the Ni(111)
transition-metal surface, the latter case representing a challenging
scenario for a theoretical description.
The results of our calculations, compared with available experimental
data and other theoretical investigations are rather good.
We confirm that the rare gas Xe atom is preferentially adsorbed on the
top-site configuration on the Ni(111) surface too, similarly to what
happens on other metallic surfaces.
Moreover, our approach, based on the use of the MLWFs, allow us 
to well characterize the bonds between graphene
and Ni(111) which is clearly chemical in character is spite of the 
relatively small binding energy. 
This opens the way to applications of the method to investigate 
adsorption processes also on other interesting transition-metal surfaces such 
as Rh(111). 

\section{Acknowledgements}
We thank very much R. Sabatini for help in performing rVV10 calculations
and L. Vattuone for useful discussions; we also
acknowledge financial support from MIUR (Grant PRIN 2010-2011 GRAF).

\vfill
\eject

\begin{table}
\caption{Binding energy E$_b$ (in meV) and (in parenthesis) equilibrium 
distance R (in \AA) for Xe adsorbed on Ni(111) in the 
{\it{top}} and {\it{hollow}} configurations, using different methods.
In square parenthesis the adsorption energy E$_a$ (in meV), 
which takes the Xe-Xe interaction into account (see text). 
Results are compared with available experimental reference data.}
\begin{center}
\begin{tabular}{|l|c|c|}
\hline
method            &   {\it{top}}   &   {\it{hollow}}  \\ \tableline
\hline
LDA               &   -401 (2.51)   &   -382 (2.17) \\        
PBE               &    -19 (3.93)   &    -18 (3.91) \\     
\hline
PBE+D             &   -391 (2.76)   &   -354 (3.25) \\
rVV10             &   -284 (3.33)   &   -271 (3.41) \\
vdW-DF2           &   -178 (3.80)   &   -174 (3.84) \\
\hline
DFT/vdW-WF2s1     &   -283 (2.98)   &   -229 (3.05) \\
DFT/vdW-WF2s1     &  [-369]         &  [-316] \\
\hline
 expt.$^a$    &  [-369]           &  --- \\
 expt.$^b$    &  [-317]           &  --- \\
 expt.$^c$    &  [-191 $\pm 9$]   &  --- \\
\hline
\end{tabular}
\tablenotetext[1]{ref.\onlinecite{Vidali}.}        
\tablenotetext[2]{ref.\onlinecite{Dolle}.}        
\tablenotetext[3]{ref.\onlinecite{Wong}.}        
\end{center}
\label{table1}
\end{table}
\vfill
\eject

\begin{table}
\caption{Binding energy E$_b$ (in meV per C atom, a positive value means
that the configuration is unbound) and (in parenthesis) equilibrium 
distance R (in \AA) for graphene on Ni(111), considering the {\it{top-fcc}},
{\it{bridge-top}}, and {\it{fcc-hcp}} configurations,
using different methods, compared with available
experimental and theoretical reference data.
R is defined as the separation between the average $z$ coordinate of the
C atoms of the graphene layer and that of the topmost Ni atoms of the 
Ni(111) substrate.}
\begin{center}
\begin{tabular}{|l|c|c|c|}
\hline
method              & {\it{top-fcc}} & {\it{bridge-top}} & {\it{fcc-hcp}} \\ \tableline
\hline
LDA                     &  -196 (2.01)&-210 (1.95)   &   -35 (3.24) \\
PBE                     &    +9 (2.17)&  +9 (2.09)   &    -5 (3.85) \\     
PBEsol                  &  -131 (2.08)&-138 (1.99)   &   -18 (3.49) \\     
\hline
DFT-D2$^a$              &  -160 (2.09)&-157 (2.05)   & -89 (3.02)   \\
DFT-D3$^b$              &  -107 (3.25)&   ---        &   ---        \\
rVV10                   &   -81 (3.14)& -80 (3.11)   &   -75 (3.44) \\     
vdW-DF$^c$              &   -37 (3.50)&   ---        &   ---        \\
vdW-DF2$^d$             &   -44 (3.68)&   ---        &   ---        \\
C09$^d$                 &  -141 (2.07)&   ---        &   ---        \\
M06L$^e$                &   -44 (2.37)&   ---        &   ---        \\
optB86b$^f$             &  -112 (2.12)&   ---        &   ---        \\
optB88vdW$^g$           &   -71 (2.18)&   ---        &   ---        \\
RPA$^f$                 &   -67 (2.17)&   ---        &   ---        \\
RPA$^h$                 &   -70 (2.19)&   ---        &   ---        \\
\hline
DFT/vdW-WF2s1           &  -129 (2.12)&   -128 (1.99)&   -48 (3.64) \\
DFT/vdW-WF2s1-shift     &   -29 (3.84)&      ---     &   ---        \\
\hline
expt.$^i$               &       (2.11$\pm 0.07$)&   ---   &   ---   \\
\hline
\end{tabular}
\tablenotetext[1]{ref.\onlinecite{Hasegawa}.}        
\tablenotetext[2]{ref.\onlinecite{Feng}.}        
\tablenotetext[3]{ref.\onlinecite{Vanin}.}        
\tablenotetext[4]{ref.\onlinecite{Hamada}.}        
\tablenotetext[5]{ref.\onlinecite{Andersen}.}        
\tablenotetext[6]{ref.\onlinecite{Mittendorfer}.}        
\tablenotetext[7]{ref.\onlinecite{Zhang}.}        
\tablenotetext[8]{ref.\onlinecite{Olsen}.}        
\tablenotetext[9]{ref.\onlinecite{Gamo}.}     
\end{center}
\label{table2}
\end{table}
\vfill
\eject

\pagestyle{empty}

\begin{figure}
\centerline{
\includegraphics[width=19cm]{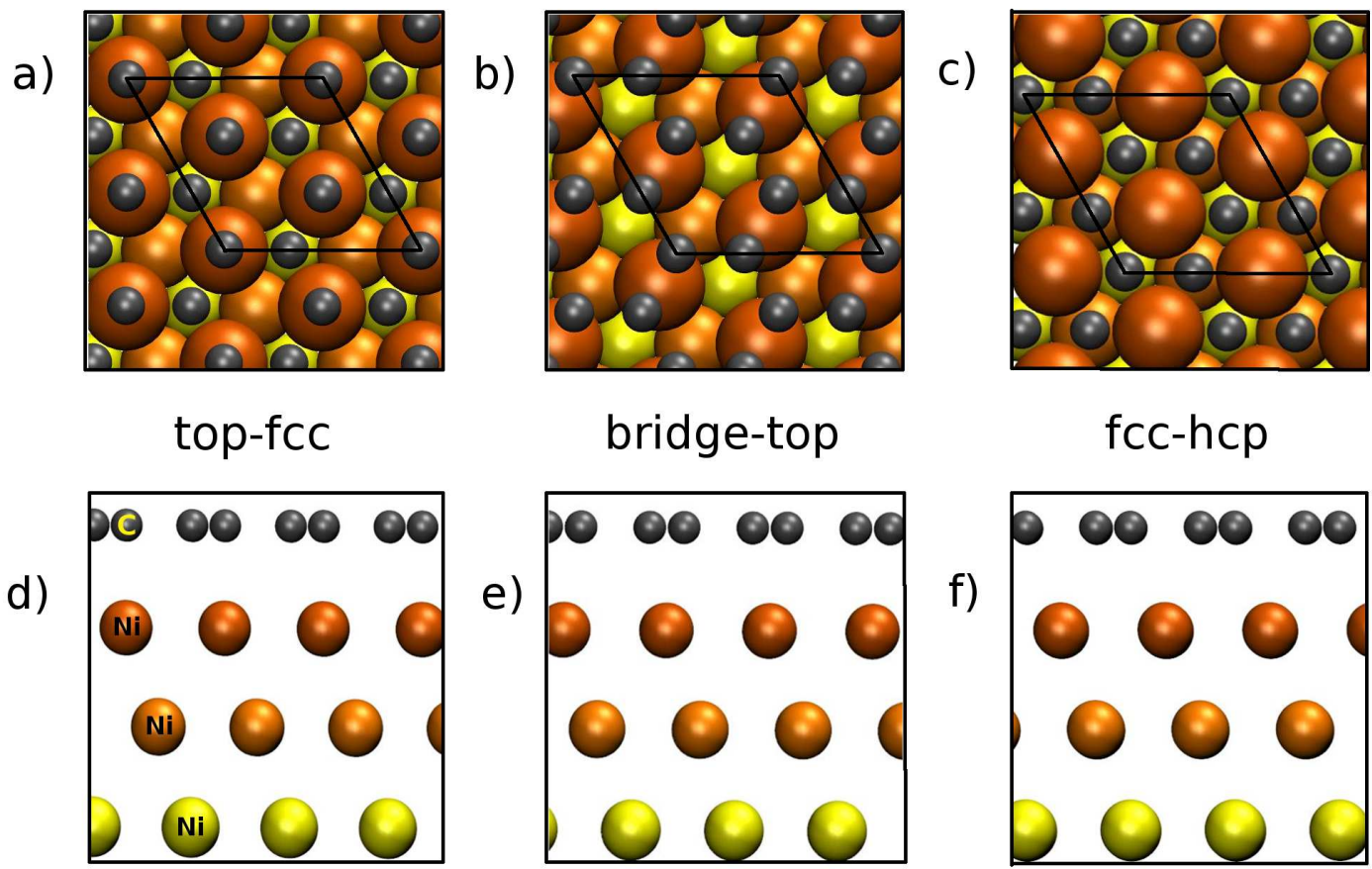}
}
\caption{Plane (upper panels) and side (lower panels) views of graphene
on the Ni(111) surface, in the {\it{top-fcc}}, {\it{bridge-top}}, 
and {\it{fcc-hcp}} configuration; the adopted simulation cell is shown in the
plane views.}
\label{fig1}
\huge
\end{figure}
                      
\begin{figure}
\centerline{
\hspace{1.0cm}
\includegraphics[angle=-90,width=21cm]{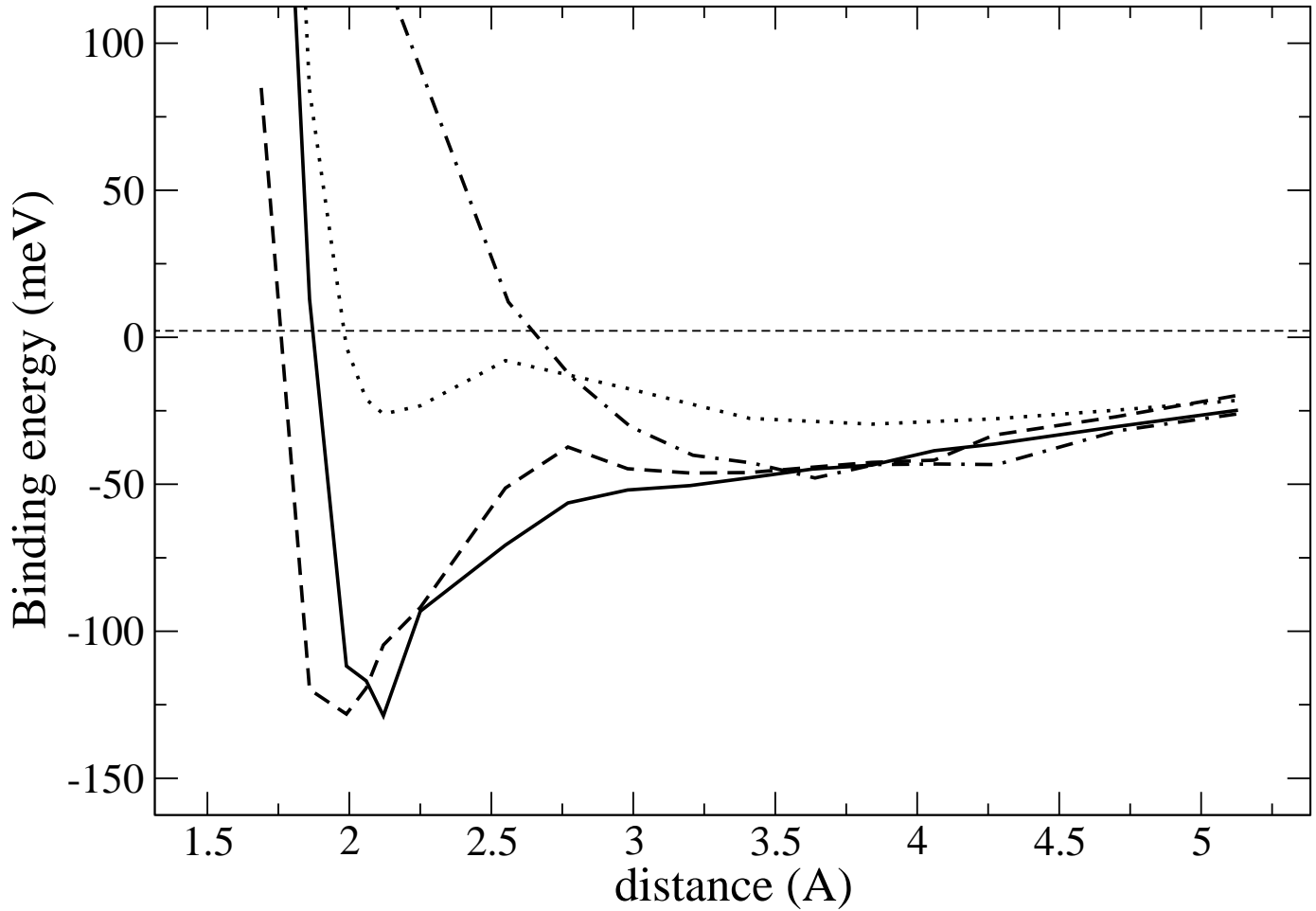}
}
\caption{Binding energy (per C atom) of graphene on Ni(111) 
computed by the DFT/vdW-WF2s1
method, in the {\it{top-fcc}} (solid line), {\it{bridge-top}} (dashed line), 
and {\it{fcc-hcp}} (dot-dashed line) configuration,
as a function of the distance between the graphene layer and 
the Ni(111) surface. The dotted line has been obtained using the
DFT/vdW-WF2s1-shift scheme (see text).}
\label{fig2}
\huge
\end{figure}
                      
\begin{figure}
\vspace{-3.3cm}
\centerline{
\includegraphics[width=20cm]{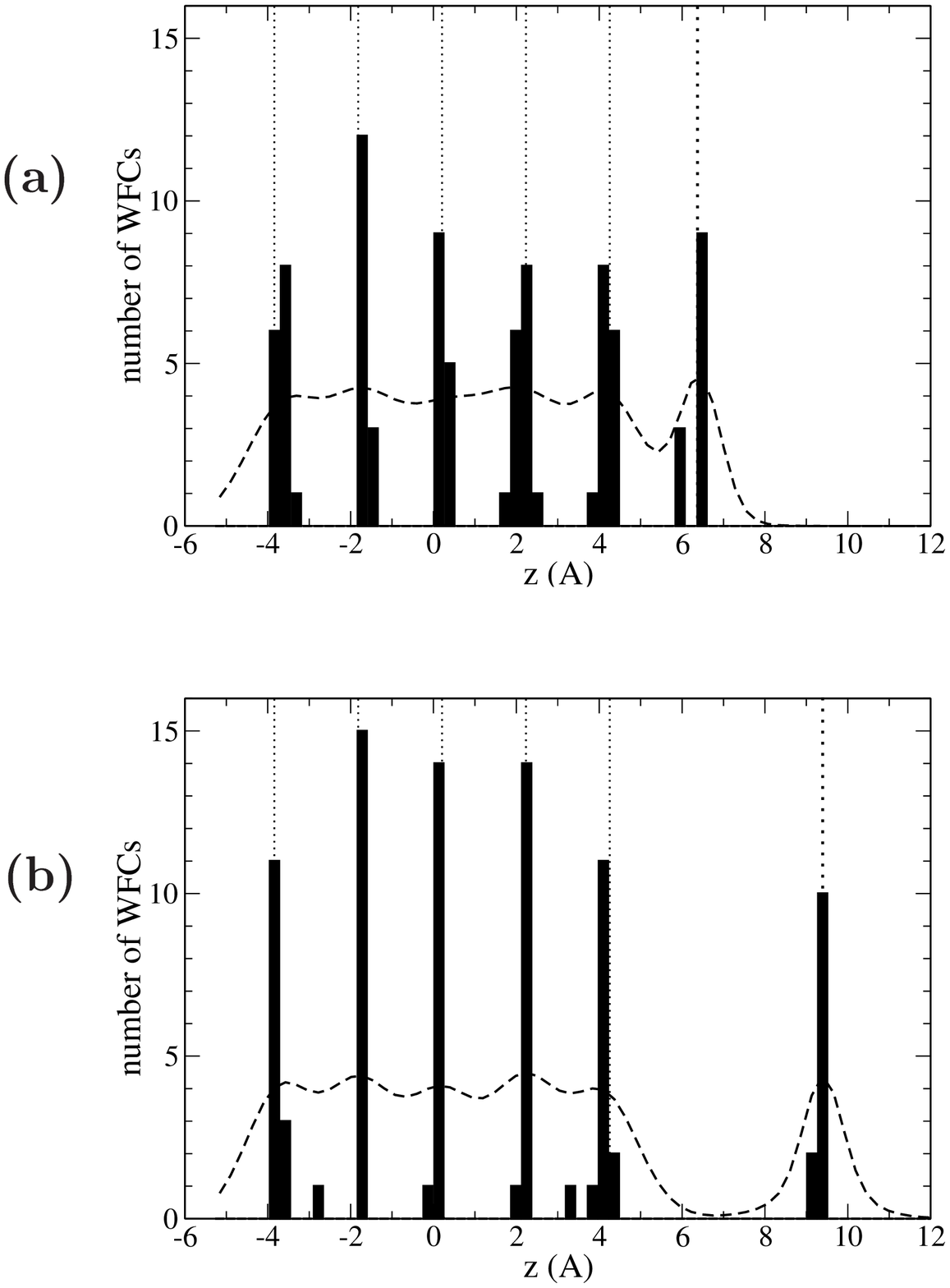}
}
\vspace{-4.5cm}
\caption{Histograms showing the distribution of the WFCs along the
$z$ coordinate (perpendicular to the Ni(111) surface) for the 
{\it{top-fcc}} configuration for graphene on Ni(111):
(a) corresponds to the binding energy minimum at the 2.12 \AA\ distance,
while (b) corresponds to the 5.13 \AA\ distance. 
The dashed curve has been obtained by a Gaussian smearing where the
Gaussian widths are just given by the spreads of the MLWFs, while
the vertical dotted lines denote the average position of the Ni atoms 
belonging to the different Ni(111) layers and (thicker line on the right) 
of the C atoms of graphene.}
\label{fig3}
\huge
\end{figure}


\begin{thebibliography}{9}
\bibitem{Wintterlin} J. Wintterlin and M. L. Bocquet, 
                     Surf. Sci. {\bf 603}, 1841 (2009).
\bibitem{Voloshina2012} E. Voloshina and Y. Dedkov,
                        Phys. Chem. Chem. Phys. {\bf 14}, 13502 (2012).
\bibitem{Batzill} M. Batzill,
                  Surf. Sci. Rep. {\bf 67}, 83 (2012).
\bibitem{Nagashima} A. Nagashima, N. Tejima, C. Oshima,
                    Phys. Rev. B {\bf 50}, 17487 (1994).
\bibitem{Gamo} Y. Gamo, A. Nagashima, M. Wakabayashi, M. Terai, and C. Oshima,
               Surf. Sci. {\bf 374}, 61 (1997).
\bibitem{Fuentes} M. Fuentes-Cabrera, M. I. Baskes, A. V. Melechko, and 
                  M. L. Simpson, Phys. Rev. B {\bf 77}, 035405 (2008). 
\bibitem{Zhao} W. Zhao, S. M. Kozlov, O. H\"ofert, K. Gotterbarm, 
               M. P. A. Lorenz, F. Vi\~{n}es, C. Papp, A. G\"orling, 
               and H.-P. Steinr\"uck, 
               Phys. Chem. Lett. {\bf 2}, 759 (2011).
\bibitem{DedkovPRL} Y. S. Dedkov, M. Fonin, U. Rudiger, and C. Laubschat, 
                    Phys. Rev. Lett. {\bf 100}, 107602 (2008).
\bibitem{Gruneis} A. Gr\"uneis, K. Kummer, and D. V. Vyalikh, 
                  New J. Phys. {\bf 11}, 073050 (2009).
\bibitem{Dedkov} Yu. S. Dedkov and M. Fonin, 
                 New J. Phys. {\bf 12}, 125004 (2010).
\bibitem{Li} M. Li, J. B. Hannon, R. M. Tromp, J. Sun, J. Li, 
             V. B. Shenoy, and E. Chason, 
             Phys. Rev. B {\bf 88}, 041402(R) (2013).
\bibitem{Allard} A. Allard and L. Wirtz,
                 Nano Lett. {\bf 10}, 4335 (2010).
\bibitem{Riccardi} P. Riccardi, A. Cupolillo, M. Pisarra, A. Sindona, 
                   and L. S. Caputi, 
                   Appl. Phys. Lett. {\bf 97}, 221909 (2010).
\bibitem{DedkovJAP} Y. S. Dedkov, M. Sicot, and M. Fonin,
                   J. Appl. Phys. {\bf 107}, 09E121 (2010).§
\bibitem{Cho} Y. Cho, Y. C. Choi, and K. S. Kim,
              J. Phys. Chem. C {\bf 115}, 6019 (2011).
\bibitem{Odahara} G. Odahara, S. Otani, C. Oshima, M. Suzuki, T. Yasue, 
                  and T. Koshikawa, Surf. Sci. {\bf 605}, 1095 (2011).
\bibitem{Garcia} A. Garcia-Lekue, T. Balashov, M. Olle, G. Ceballos, 
                 A. Arnau, P. Gambardella, D. Sanchez-Portal, and A. Mugarza,
                 Phys. Rev. Lett. {\bf 112}, 066802 (2014).
\bibitem{Bianchini} F. Bianchini, L. L. Patera, M. Peressi, C. Africh, 
                    and G. Comelli, J. Phys. Chem. Lett. {\bf 5}, 467 (2014).
\bibitem{Zhang} W. B. Zhang, C. Chen, P.-Y. Tang,
                J. Chem. Phys. {\bf 141}, 044708 (2014).
\bibitem{Mittendorfer} F. Mittendorfer, A. Garhofer, J. Redinger, 
                       J. Klime\v{s}, J. Harl, G. Kresse,
                       Phys. Rev. B {\bf 84}, 201401(R) (2011).
\bibitem{Parreiras} D. E. Parreiras, E. A. Soares, G. J. P. Abreu,
                    T. E. P. Bueno, W. P. Fernandes, V. E. de Carvalho, 
                    S. S. Carara, H. Chacham, and R. Paniago, 
                    Phys. Rev. B {\bf 90}, 155454 (2014).
\bibitem{Sun} X. Sun, S. Entani, Y. Yamauchi, A. Pratt, and M. Kurahashi, 
              J. Appl. Phys. {\bf 114}, 143713 (2013).
\bibitem{Hamada} I. Hamada, M. Otani,
                 Phys. Rev. B {\bf 82}, 153412 (2010).
\bibitem{Olsen} T. Olsen, K. S. Thygesen, 
                Phys. Rev. B {\bf 87}, 075111 (2013).
\bibitem{Feng} X. Li, J. Feng, E. Wang, S. Meng, J. Klimes, A. Michaelides,
               Phys. Rev. B {\bf 85}, 085425 (2012).
\bibitem{Vanin} M. Vanin, J.J. Mortensen, A.K. Kelkkanen, J.M. Garcia-Lastra,
                K.S. Thygesen, and K.W. Jacobsen,
                Phys. Rev. B {\bf 81}, 081408(R) (2010).
\bibitem{Andersen} M. Andersen, L. Hornekaer, B. Hammer,
                   Phys. Rev. B {\bf 86}, 085405 (2012).
\bibitem{Karpan} V. M. Karpan, P. A. Khomyakov, A. A. Starikov, 
                 G. Giovannetti, M. Zwierzycki, M. Talanana, G. Brocks, 
                 J. van den Brink, and P. J. Kelly, 
                 Phys. Rev. B {\bf 78}, 195419 (2008).
\bibitem{OlsenPRL} T. Olsen, K. S. Thygesen, 
                   Phys. Rev. Lett. {\bf 112}, 203001 (2014).
\bibitem{Ren} X. Ren, P. Rinke, C. Joas, M. Scheffler, 
              J Mater Sci {\bf 47}, 7447 (2012).
\bibitem{Harl} J. Harl and G. Kresse, 
               Phys. Rev. Lett. {\bf 103}, 056401 (2009).
\bibitem{Sutter} P. Sutter, J. T. Sadowski, and E. Sutter,
                 Phys. Rev. B {\bf 80}, 245411 (2009).
\bibitem{PBE+D} S. Grimme, 
                J. Comp. Chem. {\bf 27}, 1787 (2006);
                V. Barone, M. Casarin, D. Forrer, M. Pavone, M. Sambi, 
                A. Vittadini,
                J. Comp. Chem. {\bf 30}, 934 (2009).
\bibitem{Kozlov} S. M. Kozlov, F. Vies, and A. G\"orling, 
                 J. Phys. Chem. C {\bf 116}, 7360 (2012).
\bibitem{Sun11} X. Sun, Y. Yamauchi, 
                J. Appl. Phys. {\bf 110}, 103701 (2011).
\bibitem{Hasegawa} M. Hasegawa, K. Nishidate, T. Hosokai, and N. Yoshimoto, 
                   Phys. Rev. B {\bf 87}, 085439 (2013).
\bibitem{Riley}  K. E. Riley, M. Pito\v{n}\'{a}k, P. Jure\v{c}ka, 
                 P. Hobza,  Chem. Rev. {\bf 110}, 5023 (2010). 
\bibitem{MRS} A. Tkatchenko, L. Romaner, O. T. Hofmann, E. Zojer, 
              C. Ambrosch-Draxl, and M. Scheffler, 
              MRS Bulletin, {\bf 35}, 435 (2010).
\bibitem{Klimes} J. Klime\v{s}, A. Michaelides, 
                 J. Chem. Phys. {\bf 137}, 120901 (2012).
\bibitem{Marzari} N. Marzari and D. Vanderbilt,
                  Phys. Rev. B {\bf 56}, 12847 (1997).
\bibitem{silvprl}  P. L. Silvestrelli,  
                   Phys. Rev. Lett {\bf 100}, 053002 (2008).
\bibitem{silvmetodo} P. L. Silvestrelli,  
                     J. Phys. Chem. A {\bf 113}, 5224 (2009).
\bibitem{silvsurf} P. L. Silvestrelli, K. Benyahia, S. Grubisi\^{c}, 
                   F.  Ancilotto, F. Toigo,  
                   J. Chem. Phys. {\bf 130}, 074702 (2009).
\bibitem{CPL} P. L. Silvestrelli,
              Chem. Phys. Lett. {\bf 475}, 285 (2009). 
\bibitem{silvinter} P. L. Silvestrelli, F.  Toigo, F.  Ancilotto, 
                    J. Phys. Chem. C {\bf 113}, 17124 (2009).
\bibitem{ambrosetti} A. Ambrosetti, P. L. Silvestrelli, 
                     J. Phys. Chem. C {\bf 115}, 3695 (2011).
\bibitem{Costanzo} F. Costanzo, P. L. Silvestrelli, Francesco Ancilotto,
                   J. Chem. Theory Comp. {\bf 8}, 1288 (2012);
                   Archives of Metallurgy and Materials {\bf 57}, 1075 (2012). 
\bibitem{Ar-Pb} P. L. Silvestrelli, A. Ambrosetti, S. Grubisi\^{c}, 
                and F. Ancilotto,
                Phys. Rev. B {\bf 85}, 165405 (2012).
\bibitem{Ambrosetti2012} A. Ambrosetti, F. Ancilotto, P. L. Silvestrelli,
                         J. Phys. Chem. C {\bf 117}, 321 (2013).
\bibitem{C3} A. Ambrosetti, P. L. Silvestrelli,
             Phys. Rev. B {\bf 85}, 073101 (2012).
\bibitem{PRB2013} P. L. Silvestrelli and A. Ambrosetti,
                  Phys. Rev. B {\bf 87}, 075401 (2013).
\bibitem{QHO-WF} P. L. Silvestrelli,
                 J. Chem. Phys. {\bf 139}, 054106 (2013).
\bibitem{QHO-WFs} P. L. Silvestrelli, A. Ambrosetti,
                  J. Chem. Phys. {\bf 140}, 124107 (2014).
\bibitem{Dion}  M. Dion, H. Rydberg, E. Schr\"oder, D. C. Langreth,
                B. I. Lundqvist,  Phys. Rev. Lett.
                {\bf 92}, 246401 (2004);
                G. Roman-Perez, J. M. Soler,
                Phys. Rev. Lett. {\bf 103}, 096102 (2009).
\bibitem{Langreth07} T. Thonhauser, V. R. Cooper, S. Li,  
                     A. Puzder, P. Hyldgaard, D. C. Langreth, 
                     Phys. Rev. B {\bf 76}, 125112 (2007).
\bibitem{Lee-bis} K. Lee, \'E. D. Murray, L. Kong, B. I. Lundqvist, and 
                  D. C. Langreth, 
                  Phys. Rev. B {\bf 82}, 081101(R) (2010).
\bibitem{Sabatini} R. Sabatini, T. Gorni, S. de Gironcoli,
                   Phys. Rev. B {\bf 87}, 041108(R) (2013).
\bibitem{PBE} J. P. Perdew, K. Burke, M. Ernzerhof,
              Phys. Rev. Lett. {\bf 77}, 3865 (1996).
\bibitem{london} R. Eisenhitz and F. London, Z. Phys. {\bf 60}, 491 (1930).
\bibitem{polvol} T. Brink, J. S. Murray, P. Politzer, 
                 J. Chem. Phys. {\bf 98}, 4305 (1993).
\bibitem{Bondi} A. Bondi, 
                J. Phys. Chem. {\bf 68}, 441 (1964).
\bibitem{Grimme} S. Grimme, J. Antony, T. Schwabe, C. M\"uck-Lichtenfeld, 
                 Org. Biomol. Chem.  {\bf 5}, 741 (2007);
                 S. Grimme, J. Antony, S. Ehrlich, H. Krieg,
                 J. Chem. Phys. {\bf 132}, 154104 (2010).
\bibitem{ESPRESSO} See www.quantum-espresso.org for information about the 
                   Quantum-ESPRESSO package.
\bibitem{WanT} See www.wannier-transport.org for information about the
               WanT package;
               see also A. Calzolari, N. Marzari, I. Souza and 
               M. Buongiorno Nardelli, Phys. Rev. B {\bf 69}, 035108 (2004).
\bibitem{Seyller} Th. Seyller, M. Caragiu, R. D. Diehl, 
                  P. Kaukasoina, M. Lindroos,
                  Chem. Phys. Lett. {\bf 291}, 567 (1998);
                  M. Caragiu, Th. Seyller, R. D. Diehl,
                  Phys. Rev. B {\bf 66}, 195411 (2002).
\bibitem{Silva} J. L. F. Da Silva, C. Stampfl, M. Scheffler,
                Phys. Rev. Lett. {\bf 90}, 066104 (2003). 
\bibitem{DaSilva05} J. L. F. Da Silva, C. Stampfl, M. Scheffler,
                  Phys. Rev. B {\bf 72}, 075424 (2005).
\bibitem{DaSilva} J. L. F. Da Silva, C. Stampfl,
                  Phys. Rev. B {\bf 77}, 045401 (2008).
\bibitem{Righi} M. C. Righi, M. Ferrario, 
                  J. Phys.: Condens. Matter {\bf 19}, 305008 (2007).
\bibitem{Lazic} P. Lazi\'{c}, \v{Z}. Crljen, R. Brako, B. Gumhalter,
                  Phys. Rev. B {\bf 72}, 245407 (2005).
\bibitem{Zhang11} Y. N. Zhang, F. Hanke, V. Bortolani, M. Persson, 
                  R. Q. Wu, 
                  Phys. Rev. Lett. {\bf 106}, 236103 (2011).
\bibitem{Abad} E. Abad, Y. J. Dappe,  J. I.  Mart\'inez, F. Flores, 
               J. Ortega, 
               J. Chem. Phys. {\bf 134}, 044701 (2011).
\bibitem{Fajin} J. L. Faj\'in, F. Illas, J. R. B. Gomes,
                J. Chem. Phys. {\bf 130}, 224702 (2009).
\bibitem{Hamada12} I. Hamada,
                   Phys. Rev. B {\bf 86}, 195436 (2012).
\bibitem{Perrot} F. Perrot, M. Rasolt,
                 J. Phys.: Condens. Matter {\bf 6}, 1473 (1994).
\bibitem{Ashcroft} See, for instance: N. W. Ashcroft and N. D. Mermin,
                   Solid State Physics, Holt-Saunders International Editions,
                   Philadelphia, 1976.
\bibitem{Diehl} R. D. Diehl, Th.  Seyller, M. Caragiu, G. S. Leatherman, 
                N. Ferralis, K. Pussi, P.  Kaukasoina, M. Lindroos,
                J. Phys.: Condens. Matter {\bf 16}, S2839 (2004).
\bibitem{Betancourt} A. E. Betancourt, D. M. Bird, 
                  J. Phys.: Condens. Matter {\bf 12}, 7077 (2000).
\bibitem{Bagus} P. S. Bagus, V. Staemmler, C. W\"oll, 
                Phys. Rev. Lett. {\bf 89}, 096104 (2002).
\bibitem{Vidali} G. Vidali, G. Ihm, H. Y. Kim, M. W. Cole,
                 Surf. Sci. Rep. {\bf 12}, 135 (1991).
\bibitem{Dolle} P. Dolle, M. Alnot, J. J. Ehrhandt, A. Thomy, A. Cassuto,
                Surf. Sci. {\bf 152/153}, 620 (1985); 
                D. Fargues, P. Dolle, M. Alnot, J. J. Ehrardt, 
                Surf. Sci. {\bf 214}, 187 (1988).
\bibitem{Wong} A. Wong, X. D. Zhu, 
               Appl. Phys. A {\bf 63}, 1 (1996).
\bibitem{Muller} J. E. M\"uller,
                 Phys. Rev. Lett. {\bf 65}, 3021 (1990).
\bibitem{PBEsol} J. P. Perdew, A. Ruzsinszky, G. I. Csonka, O. A. Vydrov,
                 G. E. Scuseria, L. A. Constantin, X. Zhou, K. Burke,
                 Phys. Rev. Lett. {\bf 100}, 136406 (2008).
\bibitem{Liu} W. Liu, J. Carrasco, B. Santra, A. Michaelides, M. Scheffler,
              A. Tkatchenko,
              Phys. Rev. B {\bf 86}, 245405 (2012).
\end{thebibliography}
\end{document}